\documentclass[runningheads]{llncs}

\usepackage{amssymb}
\usepackage{graphicx}
\usepackage{amsmath}
\usepackage{url}
\usepackage{times}
\usepackage{enumitem}
\usepackage{mymacros}

\usepackage{url} 
\usepackage{graphicx}

\def\UnivA{{\mathcal{A}}}
\def\UnivC{{\mathcal{C}}}
\def\UnivT{{\mathcal{T}}}
\def\UnivN{{\mathcal{N}}}
\def\UnivV{{\mathcal{V}}}
\def\UnivM{{\mathcal{M}}}
\def\UnivE{{\mathcal{E}}}

\def\beginproof{\begin{proof}\vspace{-0.1cm}}

\def\hide#1{}

\begin{document}

\title{May I Take Your Order?}
\subtitle{On the Interplay Between Time and Order in Process Mining}
\titlerunning{May I Take Your Order?}

\author{Wil M.P. van der Aalst\inst{1,2} \and Luis Santos\inst{1}}

\institute{
Process and Data Science (Informatik 9), RWTH Aachen University, Aachen, Germany \and
Fraunhofer-Institut f\"{u}r Angewandte Informationstechnik (FIT), Sankt Augustin, Germany \\
\email{wvdaalst@pads.rwth-aachen.de}}
\maketitle

\begin{abstract}
Process mining starts from event data. 
The ordering of events is vital for the discovery of process models.
However, the timestamps of events may be unreliable or imprecise.
To further complicate matters, also causally unrelated events may be ordered in time.
The fact that one event is followed by another does not imply that the former causes the latter.
This paper explores the relationship between time and order.
Moreover, it describes an approach to preprocess event data having timestamp-related problems.
This approach avoids using accidental or unreliable orders and timestamps, 
creates partial orders to capture uncertainty, and allows for exploiting domain knowledge to (re)order events. 
Optionally, the approach also generates interleavings to be able to use existing process mining techniques 
that cannot handle partially ordered event data.
The approach has been implemented using ProM and can be applied to any event log.

\end{abstract}

\setcounter{footnote}{0}

\section{Introduction}
\label{sec:intro}

Most process mining techniques require the events within a case to the totally ordered \cite{process-mining-book-2016}.
For example, nearly all discovery techniques convert the event log into a multiset of traces where each trace is a sequence of activities. 
To order the events within a case, typically timestamps are used.
However, the timestamps may be unreliable or too coarse-grained. 
Consider, for example, a nurse taking a blood sample from a patient at 16.55 
but recording this into the hospital's information system at 17.55 when her shift ends (event $e_1$). At 17.15, the patient's insurance company approved the operation and this was automatically recorded (event $e_2$). The same patient also had an X-ray in the evening, but only the date is recorded (event $e_3$). 
In this example, the real ordering of events was $\langle e_1,e_2,e_3 \rangle$, but in the event log they may appear as $\langle e_3,e_2,e_1 \rangle$. Event $e_1$ happened before $e_2$ but was recorded one hour later. Event $e_3$ was the last event, but because only the date was recorded, it appeared to have happened at time 00.00. Moreover, events $e_1$ and $e_2$ were fully unrelated, so why consider the temporal order? The approval was triggered by a request submitted two days before. 
Healthcare data are notorious for having data quality problems \cite{healthcare-mining-book2015}.
However, such issues can be found in any domain 
\cite{andrews-data-quality-2020,niels-hc-data-quality-2021,suriadi-data-quality-2021}.
\begin{figure}[tb!]
{
\centering
\includegraphics[width=12cm]{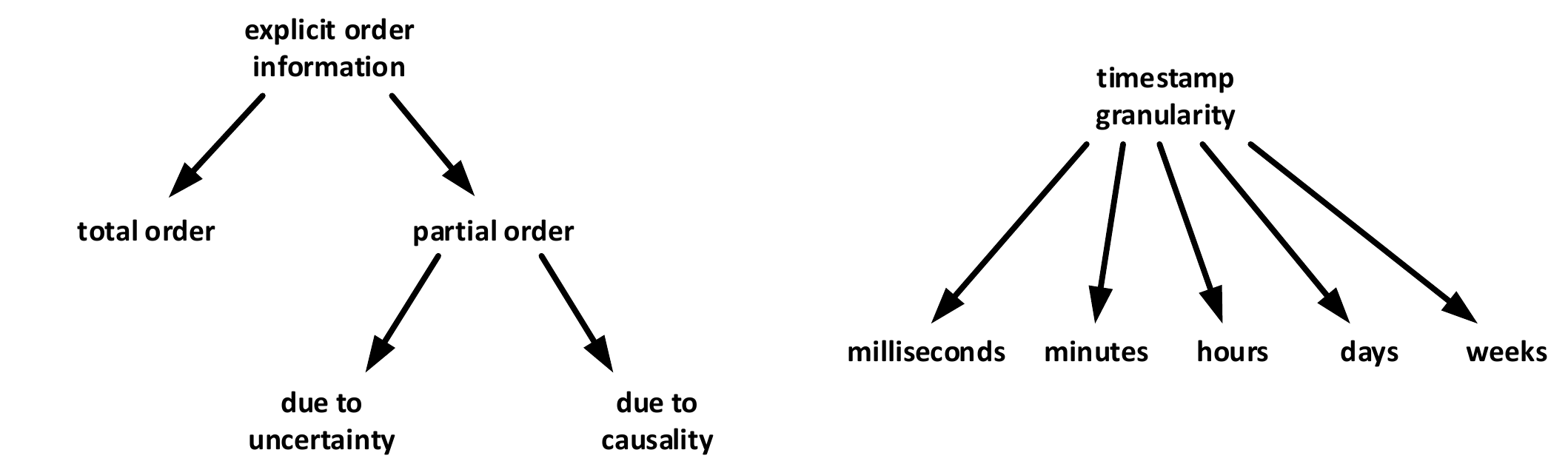}
\caption{We assume that events may have explicit order information (left) and have a timestamp (right). However, the ordering is partial and the timestamps can be coarse-grained.}\label{fig-overview}
}
\end{figure}

In this paper, we assume that events are \emph{partially ordered} and have a \emph{timestamp} (see Figure~\ref{fig-overview}).
This allows us to reason about the problems just mentioned.
Given a set of events $E$, we assume a strict partial order $\prec_o\, \subseteq E \times E$.
$e_1 \prec_o e_2$ means that event $e_1$ is before event $e_2$.
$\pi_{\mi{time}}(e_1)$ and $\pi_{\mi{time}}(e_2)$ are the timestamps of both events.
We assume that events are recorded at a certain granularity, e.g., milliseconds, seconds, hours, days, weeks, months, or years.
Events may have more fine-grained timestamps, but we map these onto the chosen level of granularity.
For example, ``19-05-2021:17.15.00'' and ``19-05-2021:17.55'' are both mapped onto ``19-05-2021'' when using days as a granularity. 

As mentioned, next to timestamps at a selected granularity level, we also assume a partial order on events.
Such a partial order can also be more coarse-grained or more fine-grained. 
One extreme is that the events are totally ordered, i.e., for any two events $e_1$ and $e_2$: $e_1 \prec_o e_2$ or $e_1 \succ_o e_2$. Another extreme is that that no two events are ordered ($e_1 \nprec_o e_2$ or $e_1 \nsucc_o e_2$).
The latter case (events are unordered) is similar to assuming that all events have the same coarse-grained timestamp (e.g., same year). 
\begin{figure}[b!]
{
\centering
\includegraphics[width=11cm]{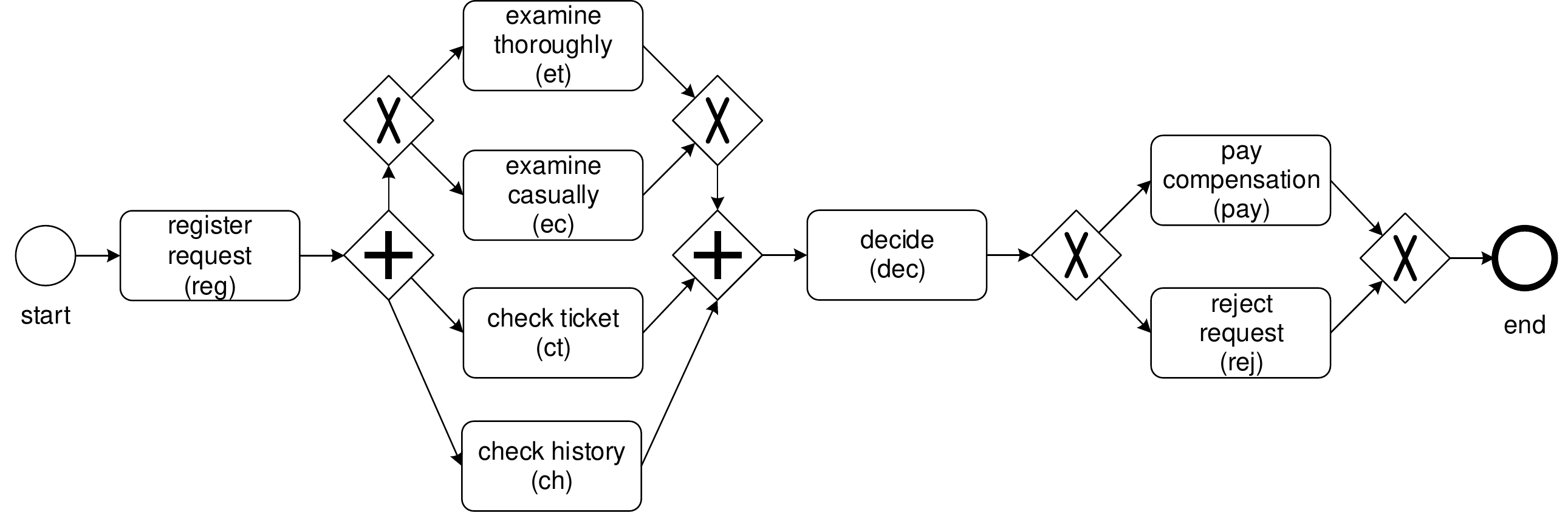}
\caption{A BPMN model having $4$ partially-ordered runs and $2 \times 2 \times 3! = 24$ sequential runs.}\label{fig-initial-bpmn-example}
}
\end{figure}

To better explain the problem, 
we consider the BPMN (Business Process Model and Notation) model 
shown in Figure~\ref{fig-initial-bpmn-example} for handling requests for compensation 
within an airline.
Customers may request compensation for various reasons, e.g., a delayed or canceled flight.
The process starts by registering the request ($\mi{reg}$).
After this, three checks need to be performed. The customer's ticket and history are checked for all cases, i.e., activities ($\mi{ct}$) and ($\mi{ch}$) need to be performed for all requests.
There is a choice between a thorough examination ($\mi{et}$) and a casual examination ($\mi{ec}$).
After this, a decision is made ($\mi{dec}$) and the request is rejected ($\mi{rej}$) or some compensation is paid ($\mi{pay}$). 
Examples of sequential runs are 
$\langle \mi{reg},\allowbreak \mi{et},\allowbreak \mi{ct},\allowbreak \mi{ch},\allowbreak \mi{dec}, \allowbreak\mi{pay} \rangle$,
$\langle \mi{reg},\allowbreak \mi{ct},\allowbreak \mi{ec},\allowbreak \mi{ch},\allowbreak \mi{dec}, \allowbreak\mi{rej} \rangle$, and
$\langle \mi{reg},\allowbreak \mi{ch},\allowbreak \mi{ct}, \allowbreak \mi{et},\allowbreak \mi{dec}, \allowbreak\mi{pay} \rangle$.
In total there are $2 \times 2 \times 3! = 24$ sequential runs.
Note that in each run there are three concurrent activities ($\mi{ct}$, $\mi{ch}$, and either $\mi{et}$ or $\mi{ec}$). These can be interleaved in $3!=6$ ways.
\begin{figure}[tb!]
{
\centering
\includegraphics[width=12cm]{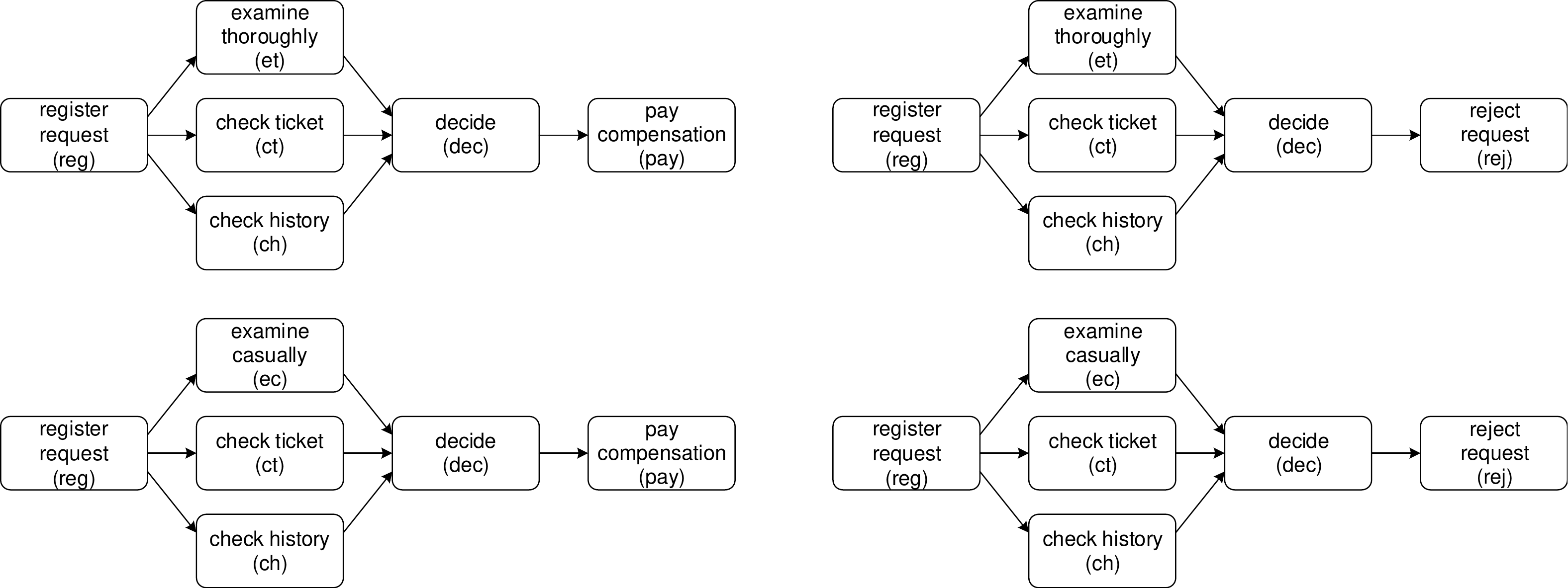}
\caption{The four partially-ordered runs of the BPMN model in Figure~\ref{fig-initial-bpmn-example}.}\label{fig-partial-order}
}
\end{figure}

There are only $2 \times 2 = 4$ partially-ordered runs. These are depicted in Figure~\ref{fig-partial-order}.
Note that the four partially-ordered runs do not need to specify the ordering of concurrent activities.
Consider the scenario with 10 concurrent activities; there is just one partially-ordered run, but there are
 $10! = 3628800$ sequential runs. Figure~\ref{fig-partial-order} helps to understand why partial orders are considered in process mining and many other analysis approaches.

\begin{table}[htbp]
\caption{A fragment of an event log where the timestamps of events are problematic.}\label{tablog}
\centering
\begin{tabular}{ccclccc}
\hline\noalign{\smallskip}
~~ event id ~~& \multicolumn{6}{c}{properties} \\ \noalign{\smallskip}\cline{2-7}\noalign{\smallskip}
   & ~ case id ~ &  ~~~~~~~~~~~~~ activity  ~~~~~~~~~~~~ & ~~ timestamp ~~ &  ~~ resource  ~~ &  ~~ cost  ~~ &  ~~ $\ldots$  ~~ \\
\noalign{\smallskip}\hline\noalign{\smallskip}
36533 & 9901 & register request & 19-05-2021:11.02.55  & Sarah & 50 & $\ldots$ \\
36534 & 9901 & check ticket & 19-05-2021:13.02  & John & 25 & $\ldots$ \\
36535 & 9902 & register request & 19-05-2021:13.02 & Sarah & 50 & $\ldots$ \\
36536 & 9902 & check history & 20-05-2021:00.00.00  & Pete & 45 & $\ldots$ \\
36537 & 9901 & check history & 20-05-2021:00.00.00  & Pete & 45 & $\ldots$ \\
36538 & 9901 & examine casually & 20-05-2021:08.55.34  & Mary & 55 & $\ldots$ \\
36539 & 9902 & check ticket & 20-05-2021:09.11.21  & John & 25 & $\ldots$ \\
36540 & 9902 & examine thoroughly & 20-05-2021:10.55  & Harry & 55 & $\ldots$ \\
36541 & 9901 & decide & 21-05-2021 & Angela & 55 & $\ldots$ \\
36542 & 9902 & decide & 21-05-2021  & Angela & 75 & $\ldots$ \\
36543 & 9902 & reject request & 22-05-2021:14.12.45  & Sarah & 20 & $\ldots$ \\
36544 & 9901 & pay compensation & 22-05-2021:16.52.37  & Sarah & 150 & $\ldots$ \\
  $\ldots$ &  $\ldots$ & $\ldots$ & $\ldots$ & $\ldots$ & $\ldots$ & $\ldots$  \\
  \noalign{\smallskip}\hline\noalign{\smallskip}
\end{tabular}
\end{table}

Table~\ref{tablog} shows a fragment of an event log corresponding to the BPMN model in Figure~\ref{fig-initial-bpmn-example}. Process discovery techniques aim to learn a process model based on such data.
If we assume the events to be sorted based on the identifier in the first column,
then case 9901 corresponds to sequential run $\langle \mi{reg},\allowbreak \mi{ct}, \allowbreak \mi{ch},\allowbreak \mi{ec},\allowbreak \mi{dec}, \allowbreak\mi{pay} \rangle$ and 
case 9902 corresponds to sequential run $\langle \mi{reg},\allowbreak \mi{ch}, \allowbreak \mi{ct},\allowbreak \mi{et},\allowbreak \mi{dec}, \allowbreak\mi{rej} \rangle$.
However, a closer inspection of the timestamp column suggests that there several problems. Some events have a precision in seconds, others in minutes, or even days. There are also timestamps of the form ``20-05-2021:00.00.00'' which suggests that times are sometimes rounded to days.
Moreover, we may know that some timestamps show the time of recording and not the actual event.
At the same time, we may know that the registration activity ($\mi{reg}$) always happens before the check activities. 

When timestamps are unreliable or imprecise, like in Table~\ref{tablog}, we cannot use them as-is. One approach is to make the timestamps more coarse-grained (e.g., just consider the day).
This automatically leads to partially ordered traces. 
Moreover, there may be explicit information that reveals explicit causal relations. For example, when the concurrent activities do not share any information.
We may know that $\mi{ct}$, $\mi{ch}$, $\mi{ec}$, and $\mi{et}$ use only data collected in $\mi{ref}$,
but that $\mi{dec}$ uses the outcomes of the three checks.
Such causal dependencies can be derived based on data-flow analysis or explicit domain knowledge, e.g., 
a payment is always preceded by a decision.
As Figure~\ref{fig-overview} shows, partial orders can be used to express either \emph{uncertainty} or \emph{explicit causality} (i.e., partial orders have a dual interpretation). 

Next to discussing the relationship between time and order, we present a concrete preprocessing approach implemented in ProM (contained in the \emph{PartialOrderVisualizer} package that can be downloaded from \url{promtools.org}). The approach uses a \emph{time aggregator}
and \emph{tiebreaker} to create a \emph{partially-ordered} event log. 
Moreover, it is possible to create a \emph{$k$-sequentialization} of 
the partially-ordered event log to be able to apply conventional approaches.

The remainder of this paper is organized as follows. Section~\ref{sec:rw} presents related work
and Section~\ref{sec:el} provides a theoretical foundation to reason about the relationship between time and order.
Section~\ref{sec:ts} presents our preprocessing approach, followed by implementation details and an example (Section~\ref{sec:impl}).
Section~\ref{sec:concl} concludes the paper.

\section{Related Work}
\label{sec:rw}

For an overview of process mining techniques, we refer to \cite{process-mining-book-2016}. 
See \cite{conf-check-book-2018} for conformance checking and \cite{process-mining-book-lars-2020} for large-scale applications in organizations such as BMW, Uber, Siemens, EDP, ABB, Bosch, and Telekom.

Recently, many papers on data quality in process mining were published \cite{andrews-data-quality-2020,niels-hc-data-quality-2021,suriadi-data-quality-2021}. Earlier
\cite{process-mining-book-2016,healthcare-mining-book2015} already provided a framework for 
data quality issues and guidelines for logging.
Timestamp-related data quality problems are seen as one of the main roadblocks in process mining.

Explicit uncertainty is considered in the work of Pegoraro et al.\ \cite{marco-ICPM-2019,marco-bis-bgs-2020}.
Event logs are annotated with explicit uncertainty and this information is used when discovering process models or checking conformance. For example, timestamps of events have upper and lower bounds and conformance checking yields optimistic and pessimistic bounds for the actual fitness.

Partial orders are a well-studied topic in modeling checking and concurrency theory.
Partially-ordered causal runs are part of the standard true-concurrency semantics of Petri net \cite{deselrunsbpm00}. See \cite{DBLP:journals/fuin/BergenthumDML09} for an example of a synthesis technique
using partial orders as input. There are a few process mining techniques that start from partial orders, e.g.,
the conformance checking technique in \cite{Xixi-Conf-Check-bpi2014-lnbip2015} and
the process discovery technique in \cite{boudewijn_runs_ToPNoC-special-issue-PN2011}.

In \cite{dss-partial-order-resolution-2020} techniques for partial order resolution are presented. 
These aim to convert a strict weak ordering into a probability distribution over all corresponding total orders.
In \cite{time-repair-bpm2020} also the ``same-timestamp problem'' is addressed, again aiming at creating total orders.
It is impossible to list all partial-order-based approaches here. 
Moreover, the goal of this paper is not to present new conformance checking or process discovery techniques. 
Instead, we provide a framework to reason about the relation between order and time, and the corresponding challenges.

In this paper, we focus on the preprocessing of event data while using standard process mining techniques. The main contribution is a discussion on the interplay between time and ordering and a concrete preprocessing tool implemented in ProM. Obviously, our framework can be combined with existing partial-order-based techniques such as \cite{dss-partial-order-resolution-2020,DBLP:journals/fuin/BergenthumDML09,time-repair-bpm2020,boudewijn_runs_ToPNoC-special-issue-PN2011,Xixi-Conf-Check-bpi2014-lnbip2015}.

\section{On the Interplay Between Time and Order in Process Mining}
\label{sec:el}

In this section, we define event logs that may have both explicit ordering information and timestamps (possibly rounded to hours, days, or weeks).
We relate such event logs to the simplified event logs typically used as input for process discovery.

\subsection{Event Logs With Time and Order}

We first define universes for events, attribute names, values, activities, timestamps, and attribute name-value mappings.
Attribute name-value mappings will be used to assign at least a case, activity, and timestamp to each event.

\begin{definition}[Universes]
$\UnivE$ is the universe of event identifiers.
$\UnivN$ is the universe of attribute names with $\{\mi{case},\mi{act},\mi{time}\} \subseteq \UnivN$,
$\UnivV$ is the universe of attribute values,
$\UnivC \subseteq \UnivV$ is the universe of case identifiers,
$\UnivA \subseteq \UnivV$ is the universe of activity names, 
$\UnivT \subseteq \UnivV$ is the universe of totally-ordered timestamps, and
$\UnivM \subseteq \UnivN \not\rightarrow \UnivV$ is the universe of attribute name-value mappings such that
for any $m \in \UnivM$: 
$\{\mi{case},\mi{act},\mi{time}\} \subseteq \mi{dom}(m)$,
$m(\mi{case}) \in \UnivC$,
$m(\mi{act}) \in \UnivA$, and
$m(\mi{time}) \in \UnivT$.
For any $n \in \UnivN$ we write $m(n) = \bot$ if $n \not\in \mi{dom}(m)$.
\end{definition}

The properties of an event are described by an attribute name-value mapping
that provides at least a case identifier, activity name, and timestamp. 
Moreover, events may have an explicit order next to timestamp information.

\begin{definition}[Event Log]\label{def:el}
An event log $L = (E,\pi,\prec_o)$ consists of a set of events $E \subseteq \UnivE$, a mapping $\pi \in E \rightarrow \UnivM$,\footnote{We use the shorthand $\pi_n(e) = \pi(e)(n)$. Note that $\pi_{\mi{case}}(e)$, $\pi_{\mi{act}}(e)$, and $\pi_{\mi{time}}(e)$ denote the case, activity, and timestamp of an event $e \in E$.} and $\prec_o\, \subseteq E \times E$ such that $(E,\prec_o)$ is a strict partial order
(i.e., irreflexive, transitive, and asymmetric).\footnote{For any $e,e_1,e_2,e_3 \in E$:
$e \nprec_o e$ (irreflexivity),  
if  $e_1 \prec_o e_2$ and $e_2 \prec_o e_3$, then $e_1 \prec_o e_3$ (transitivity), 
and if $e_1 \prec_o e_2$, then $e_2 \nprec_o e_1$ (asymmetry).}
\end{definition}

Table~\ref{tablog} shows a fragment of a larger event log.
Consider the first event in the table: 
$e=36533$, 
$\pi_{\mi{case}}(e)=9901$, 
$\pi_{\mi{act}}(e)= \mi{register\ request}$,
$\pi_{\mi{time}}(e)=$ 19-05-2021:11.02.55,
$\pi_{\mi{resource}}(e)= \mi{Sarah}$, and
$\pi_{\mi{cost}}(e)= 50$.
Table~\ref{tablog} does not define an explicit order.
Possible interpretations are that $\prec_o\ = \emptyset$ (no order) or 
a total order based on the order in the table, i.e.,
$e_1 \prec_o e_2$ if the row corresponding to $e_1$ appears before the row corresponding to $e_2$.
However, $\prec_o$ may also be based on domain knowledge or data-flow analysis (events can only use a data value produced by an earlier event).

\begin{definition}[Notations]
Let $L = (E,\pi,\prec_o)$ be an event log.
\begin{itemize}
\item $A(L) = \{ \pi_{\mi{act}}(e) \mid e \in E \}$ are the activities in $L$,
 $C(L) = \{ \pi_{\mi{case}}(e) \mid e \in E \}$ are the cases in $L$, and
 $E \tproj_c = \{ e \in E \mid \pi_{\mi{case}}(e) = c\}$ are the events of case $c \in C(L)$.
\item $\prec_t = \{(e_1,e_2) \in E \times E \mid \pi_{\mi{time}}(e_1) < \pi_{\mi{time}}(e_2)\}$ is the strict partial order based on the timestamps,
\item $\prec_{ot}\, = \prec_o \cup \prec_t$ is the union of the strict partial orders 
$\prec_o$ and $\prec_t$.
\item If two events $e_1,e_2 \in E$ are unordered with respect to $\prec_o$ (i.e., $e_1 \nprec_o e_2$ and $e_1 \nsucc_o e_2$), we write
$e_1 \sim_o e_2$.
Similarly,  $e_1 \sim_t e_2 \Leftrightarrow e_1 \nprec_t e_2 \ \wedge e_1 \nsucc_t e_2$, and
$e_1 \sim_{ot} e_2 \Leftrightarrow e_1 \nprec_{ot} e_2 \ \wedge e_1 \nsucc_{ot} e_2$.
\end{itemize}
\end{definition}


It is easy to verify that also $(E,\prec_t)$ is a strict partial order (i.e., irreflexive, transitive, and asymmetric).  $\sim_o$, $\sim_t$, and  $\sim_{ot}$ are reflexive and symmetric by construction. Note that $e_1 \sim_t e_2$ if an only if $\pi_{\mi{time}}(e_1) = \pi_{\mi{time}}(e_2)$.

\subsection{Consistency}

The relation $\prec_{ot}$, which combines $\prec_o$ and $\prec_t$, does not need to be a strict partial order. 
For example, $e_1$ happens before $e_2$ according to $\prec_o$, 
but $e_2$ happens before $e_1$ according to $\prec_t$. Because both ordering relations disagree, $\prec_{ot}$ is not asymmetric. 
Therefore, we introduce the notion of \emph{consistency}. 

\begin{definition}[Consistent]
An event log $L = (E,\pi,\prec_o)$ is consistent if for any $e_1,e_2 \in E$: 
$e_1 \prec_o e_2$ implies $\pi_{\mi{time}}(e_1) \leq \pi_{\mi{time}}(e_2)$.
\end{definition}

This can also be formulated as follows (using transposition): 
$e_1 \nprec_o e_2$ or $e_1 \nsucc_t e_2$, for any $e_1,e_2 \in E$.
Hence, it is impossible that $e_1 \prec_o e_2$ and $e_1 \succ_t e_2$ hold at the same time.
Since both orderings are not conflicting and $\prec_t$ is also a strict weak ordering, the combination yields a strict partial order.

\begin{proposition}[Consistency Implies Strict Partial Ordering]
Let $L = (E,\pi,\prec_o)$ be an event log.
$(E,\prec_t)$ is a strict weak ordering 
(i.e., a strict partial order with negative transitivity\footnote{Recall that negative transitivity means that 
if  $e_1 \nprec_t e_2$ and $e_2 \nprec_t e_3$, then $e_1 \nprec_t e_3$. 
In a strict weak ordering, incomparability is transitive, i.e., $e_1 \sim_t e_2 \ \wedge \ e_2 \sim_t e_3 \Rightarrow e_1 \sim_t e_3$.}),
and $(E,\prec_{ot})$ is a strict partial order if $L$ is consistent.
\end{proposition}
\begin{proof}
$(E,\prec_t)$ is irreflexive, transitive, and asymmetric by construction. 
Remains to show that negative transitivity holds. Assume that
$e_1 \nprec_t e_2$ and $e_2 \nprec_t e_3$, i.e.,
$\pi_{\mi{time}}(e_1) \allowbreak \geq \allowbreak \pi_{\mi{time}}(e_2)$ and 
$\pi_{\mi{time}}(e_2) \allowbreak \geq  \allowbreak \pi_{\mi{time}}(e_3)$.
Hence, $\pi_{\mi{time}}(e_1) \geq \pi_{\mi{time}}(e_3)$, i.e.,
$e_1 \nprec_t e_3$. Therefore, $(E,\prec_t)$ is a strict weak ordering.

Next, assume that $L$ is consistent. We show that $(E,\prec_{ot})$ is a strict partial order, i.e.,
for any $e,e_1,e_2,e_3 \in E$:
$e \nprec_{ot} e$ (irreflexivity),  
if  $e_1 \prec_{ot} e_2$ and $e_2 \prec_{ot} e_3$, then $e_1 \prec_{ot} e_3$ (transitivity), 
and if $e_1 \prec_{ot} e_2$, then $e_2 \nprec_{ot} e_1$ (asymmetry).
Because $\prec_{ot}\, = \prec_o \cup \prec_t$, irreflexivity follows from
$e \nprec_{o} e$ and $e \nprec_{t} e$.
Asymmetry follows directly from consistency: It is impossible that both $e_1 \prec_o e_2$ and $e_1 \succ_t e_2$
hold, so no cycles are introduced.
Transitivity relies on the fact that that negative transitivity holds for $\prec_t$.
One can use case distinction using the following four cases.
(1) $e_1 \prec_{t} e_2 \ \wedge\ \ e_2 \prec_{t} e_3 \Rightarrow e_1 \prec_{t} e_3 \Rightarrow e_1 \prec_{ot} e_3$.
(2) $e_1 \prec_{o} e_2 \ \wedge\ \ e_2 \prec_{o} e_3 \Rightarrow e_1 \prec_{o} e_3 \Rightarrow e_1 \prec_{ot} e_3$.
(3) Assume $e_1 \prec_{t} e_2 \ \wedge\ \ e_2 \prec_{o} e_3 \ \wedge\ \ e_2 \nprec_{t} e_3$.
Using consistency, we know that $e_2 \nsucc_t e_3$, hence $e_2 \sim_t e_3$.
Since $e_1 \prec_{t} e_2$ and $e_2 \sim_t e_3$, also $e_1 \prec_{t} e_3$. 
(If $e_1 \nprec_{t} e_3$, then negative transitivity implies
$e_1 \nprec_{t} e_3 \ \wedge \ e_3 \nprec_{t} e_2 \Rightarrow e_1 \nprec_{t} e_2$ leading to a contradiction.)
Since $e_1 \prec_{t} e_3$, also $e_1 \prec_{ot} e_3$.
(4) Assume $e_1 \prec_{o} e_2 \ \wedge\ \ e_2 \prec_{t} e_3 \ \wedge\ \ e_1 \nprec_{t} e_2$.
Consistency implies $e_1 \nsucc_t e_2$, hence $e_1 \sim_t e_2$.
Since $e_1 \sim_t e_2$ and $e_2 \prec_{t} e_3$, also $e_1 \prec_{t} e_3$ and $e_1 \prec_{ot} e_3$. 
Hence, in all four cases transitivity holds, thus completing the proof.
\end{proof}
\begin{figure}[tb!]
{
\centering
\includegraphics[width=12cm]{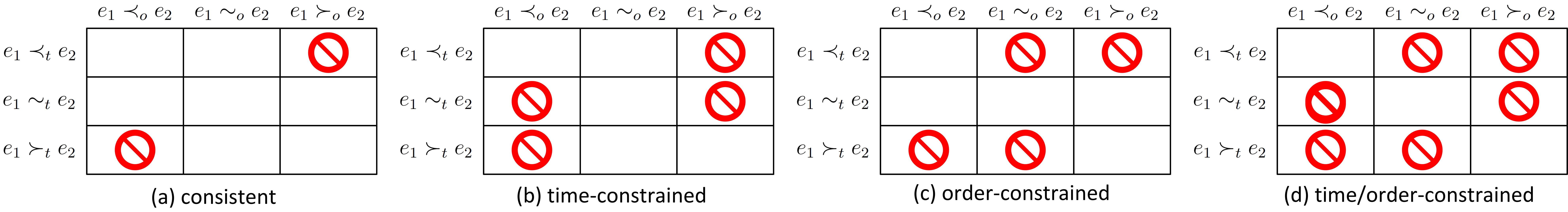}
\caption{Possible combinations of order and time relations between two events $e_1$ and $e_2$ assuming
that the event log is
(a) consistent, (b) time-constrained ($\prec_o\, \subseteq\, \prec_t$), (c) order-constrained ($\prec_t\, \subseteq\, \prec_o$), and (d) time-constrained and order-constrained ($\prec_t\, =\, \prec_o$).}\label{fig-relations}
}
\end{figure}

Both $\prec_o$ and $\prec_t$ order events.
$L$ is called time-constrained if $\prec_t$ is at least as strict as $\prec_o$, i.e.,
$e_1 \prec_o e_2$ implies $e_1 \prec_t e_2$.
$L$ is order-constrained if
$e_1 \prec_t e_2$ implies $e_1 \prec_o e_2$.
Figure~\ref{fig-relations} illustrates these notions.

\subsection{Simplified Event Logs}

The basic process discovery techniques assume linear traces and only consider the activity names. 
Therefore, we connect the more involved event log notion $L = (E,\pi,\prec_o)$ (Definition~\ref{def:el}) 
to simplified event logs and standard discovery techniques.

\begin{definition}[Simplified Event Log, Process Model, and Discovery Technique]
A trace $\sigma = \langle a_1,a_2, \ldots, a_n \rangle \in \UnivA^*$ is a sequence of activities.
$S \in \bag(\UnivA^*)$ is a simplified event log, i.e., a multiset of traces.
A process model $M \subseteq \UnivA^*$ is a set of traces.
A discovery function $\mi{disc} \in \bag(\UnivA^*) \rightarrow \powerset(\UnivA^*)$ maps an event log onto a process model.
\end{definition}

We abstract from the process model notations (e.g., BPMN or Petri nets) and focus on the modeled behavior. This allows us to define a model as a set of possible traces $M \subseteq \UnivA^*$.
$M = \mi{disc}(S)$ is the process model discovered from simplified event log $S$.
A simplified event log is a multiset of traces, e.g., $S=[\langle \mi{reg},\allowbreak \mi{ct}, \allowbreak \mi{ch},\allowbreak \mi{ec},\allowbreak \mi{dec}, \allowbreak\mi{pay} \rangle^3,\langle \mi{reg},\allowbreak \mi{ch}, \allowbreak \mi{ct},\allowbreak \mi{et},\allowbreak \mi{dec}, \allowbreak\mi{rej} \rangle^2 ]$
contains five traces.

\begin{definition}[Sequential Runs]
Let $L = (E,\pi,\prec_o)$ be a consistent event log.
For any case $c \in C(L)$, $\sigma = \langle a_1,a_2, \ldots, a_n \rangle \in \UnivA^*$ is a sequential run of $c$
if there is a bijection $f \in \{1,2, \ldots n\} \rightarrow E \tproj_c$ such that
$a_i = \pi_{\mi{act}}(f(i))$ for any $1 \leq i \leq n$ and
$e_i \nsucc_{ot} e_j$ for any $1 \leq i < j \leq n$.
$\mi{seqr}_L(c) \subseteq \UnivA^*$ are all sequential runs of case $c$.
\end{definition}

$\sigma \in \mi{seqr}_L(c)$ is a trace where each activity refers to an event of case $c$ in such a way that there is a one-to-one correspondence between the elements of $\sigma$ and $E \tproj_c$, and the order does not contradict the combined ordering $\prec_{ot}$. Given a partial order, there may be many linearizations, i.e., total orders that are compatible. In a $k$-sequentialization of $L$, we pick $k$ 
linearizations for each case. 

\begin{definition}[$k$-Sequentialization of an Event Log]\label{def:kseq}
Let $L = (E,\pi,\prec_o)$ be a consistent event log.
$S = [\sigma_1,\sigma_2, \ldots \sigma_n] \in \bag(\UnivA^*)$ is a $k$-sequentialization of $L$
if
(1) there is function $f \in \{1,2, \ldots n\} \rightarrow L(C)$ such that
$\sigma_i \in \mi{seqr}_L(f(i))$ for any $1 \leq i \leq n$, and (2)
$\card{\{ i \in \{1,2, \ldots ,n\} \mid f(i) = c\}}=k$ for any $c \in C(L)$.
$\mi{seql}_{k}(L) \subseteq \bag(\UnivA^*)$ are all possible  $k$-sequentializations of $L$.
\end{definition}

Definition~\ref{def:kseq} shows how event log $L$ can be converted into a simplified event log $S \in \mi{seql}_{k}(L)$. Each case in $L$ corresponds to $k$ linearizations in $S$. We leave it open how the linearizations are selected.
This can be probabilistic or deterministic. (In our implementation, all
linearizations are sampled from $\mi{seqr}_L(c)$ using equal probabilities).

\section{What If Timestamps Are Imprecise?}
\label{sec:ts}

As described in the introduction, timestamps may be imprecise or partially incorrect.
Therefore, we provide transformations of the event log, making time more coarse granular, e.g., all events on the same day have the same timestamp.

\begin{definition}[Time Aggregator]
$\mi{ta} \in \UnivT \rightarrow \UnivT$ is a time aggregator if for any
$t_1,t_2 \in \UnivT$ such that $t_1 < t_2$: $\mi{ta}(t_1) \leq \mi{ta}(t_2)$.
\end{definition}

For example, $\mi{ta}($19-05-2021:13.02$) = \mi{ta}($19-05-2021:17.55$) =  $19-05-2021:00.00 if a time granularity of days is used, i.e., all timestamps on the same day are mapped onto the same value.
By making time more coarse-grained, more events become unordered.
These may still be ordered by $\prec_o$, e.g., based on data-flow analysis.
Next to $\prec_o$, we may use domain knowledge in the form of a so-called \emph{tiebreaker} to optionally order events having 
identical coarse-grained timestamps.

\begin{definition}[Tiebreaker]
A tiebreaker $\prec_{tb}\, \subseteq \UnivA \times\UnivA$ is a strict partial order used to order activities having the same aggregated timestamp.
\end{definition}

A tiebreaker adds causal dependencies between events that have the same coarse-granular timestamps and belong to the same case. Using time aggregator $\mi{ta}$
and tie\-breaker $\prec_{tb}$, we can create a new event log $L^{\mi{ta},\prec_{tb}}$.

\begin{definition}[Preprocessing]
Let $L = (E,\pi,\prec_o)$ be a consistent event log,
$\mi{ta} \in \UnivT \rightarrow \UnivT$ a time aggregator, and 
$\prec_{tb}\, \subseteq \UnivA \times\UnivA$ a tiebreaker.
$L^{\mi{ta},\prec_{tb}} = (E,\pi',\prec_o')$ is the event log after applying the time aggregator $\mi{ta}$
and tiebreaker $\prec_{tb}$ such that
$\pi_n'(e) = \pi_n(e)$ for any $e \in E$ and $n \in \UnivN \setminus \{\mi{time}\}$,
$\pi_{\mi{time}}'(e) = \mi{ta}(\pi_{\mi{time}}(e))$ for any $e \in E$, and
$\prec_o' \, = \, \prec_o \cup\, \{ (e_1,e_2) \in E \times E \mid 
\pi_{\mi{case}}(e_1) = \pi_{\mi{case}}(e_2) \ \wedge \
\pi_{\mi{time}}'(e_1) = \pi_{\mi{time}}'(e_2) \ \wedge \
\pi_{\mi{act}}(e_1) \prec_{tb} \pi_{\mi{act}}(e_2)
\}$.
\end{definition}

As long as the tiebreaker $\prec_{tb}$ does not contradict $\prec_o$,
$\prec_o'$ is a partial order and $L^{\mi{ta},\prec_{tb}} = (E,\pi',\prec_o')$
is a consistent event log.

Hence, we can compute a $k$-sequentialization of $L^{\mi{ta},\prec_{tb}}$
and produce a simplified event log $S^{L,\mi{ta},\prec_{tb}} \in \mi{seql}_{k}(L^{\mi{ta},\prec_{tb}})$. 
As mentioned before, we leave it open how sequential runs are selected. 
Using the simplified preprocessed event log, we can apply any process discovery technique to obtain process model
$M^{L,\mi{ta},\prec_{tb}} = \mi{disc}(S^{L,\mi{ta},\prec_{tb}})$.

\section{Implementation}
\label{sec:impl}

The new ProM package \emph{PartialOrderVisualizer} implements the approach described in Section~\ref{sec:ts} and can be downloaded as part of ProM's nightly builds (\url{http://www.promtools.org/doku.php?id=nightly}).
It has been implemented as a so-called ``visualizer'' and can be selected by choosing \emph{Explore Partial Orders (Variants)} in the pull-down menu.
The visualizer can be applied to any event log.
\begin{figure}[tb!]
{
\centering
\includegraphics[width=12cm]{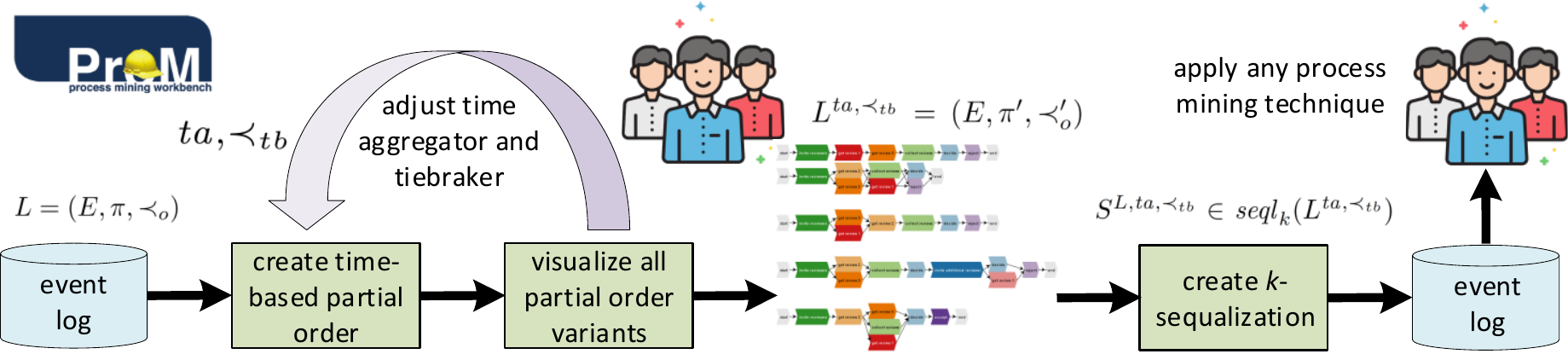}
\caption{Overview of the functionality of the \emph{PartialOrderVisualizer} package. The user can change the time granularity and modify the tiebreaker using domain knowledge. Cases with the same partial order are grouped into partial-order variants. These can be sorted and inspected. At any point in time, it is possible to create a regular event log (using $k$-sequentialization).}\label{fig-archi}
}
\end{figure}

Figure~\ref{fig-archi} shows the main components of the \emph{PartialOrderVisualizer}. 
Based on time aggregator $\mi{ta}$ and tiebreaker $\prec_{tb}$ the partial orders are computed and visualized.
Initially, $\mi{ta}$ is set to hours and $\prec_{tb} = \emptyset$. 
The user can experiment with the time granularity and add ordering constraints.
One can inspect partial order variants and details of the corresponding cases.
Moreover, the user can create a $k$-sequentialization of $L^{\mi{ta},\prec_{tb}}$. 
The resulting event log can be analyzed using classical process mining techniques.
\begin{figure}[tb!]
{
\centering
\includegraphics[width=12cm]{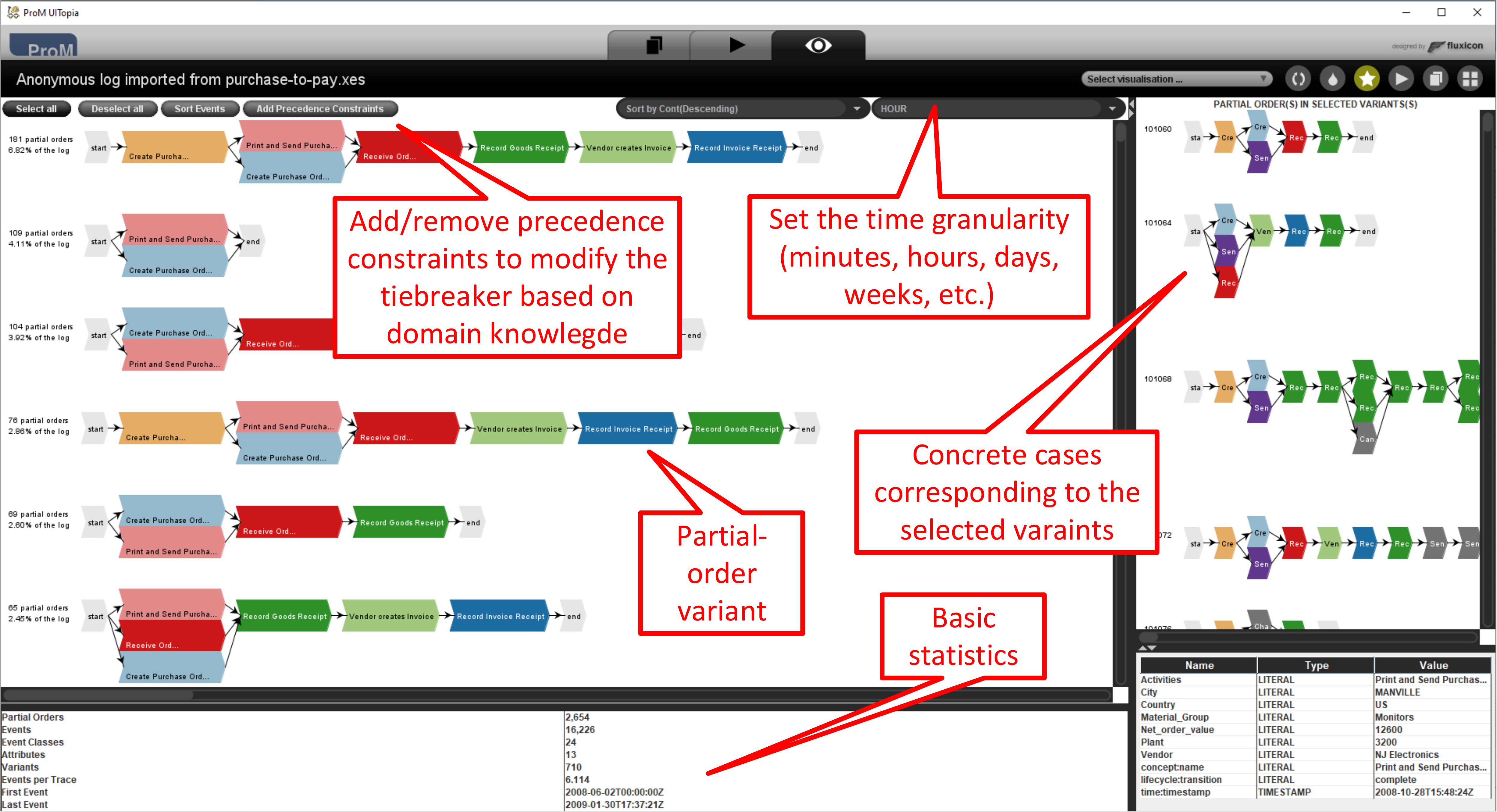}
\caption{Data from a Purchase-to-Pay (P2P) process visualized using \emph{PartialOrderVisualizer}. What can be seen is that activities \emph{Create Purchase Order Item} and \emph{Print and Send Purchase Order} often happen in the same hour.}\label{fig-tool1}
}
\end{figure}

Figure~\ref{fig-tool1} shows the \emph{PartialOrderVisualizer} for the event data of a Purchase-to-Pay (P2P) process with 2,654 cases and 16,226 events.
There are 685 trace variants in the original event log. 
The view shown in Figure~\ref{fig-tool1} uses a time granularity of one hour leading to a similar number of partial-order variants (i.e., 710).
However, one can clearly see that some activities often happen within the same hour. It is possible to add ordering information to make these sequential (if desired). Note that seeing which activities happen in the same period is valuable and provides new insights. 
\begin{figure}[tb!]
{
\centering
\includegraphics[width=12cm]{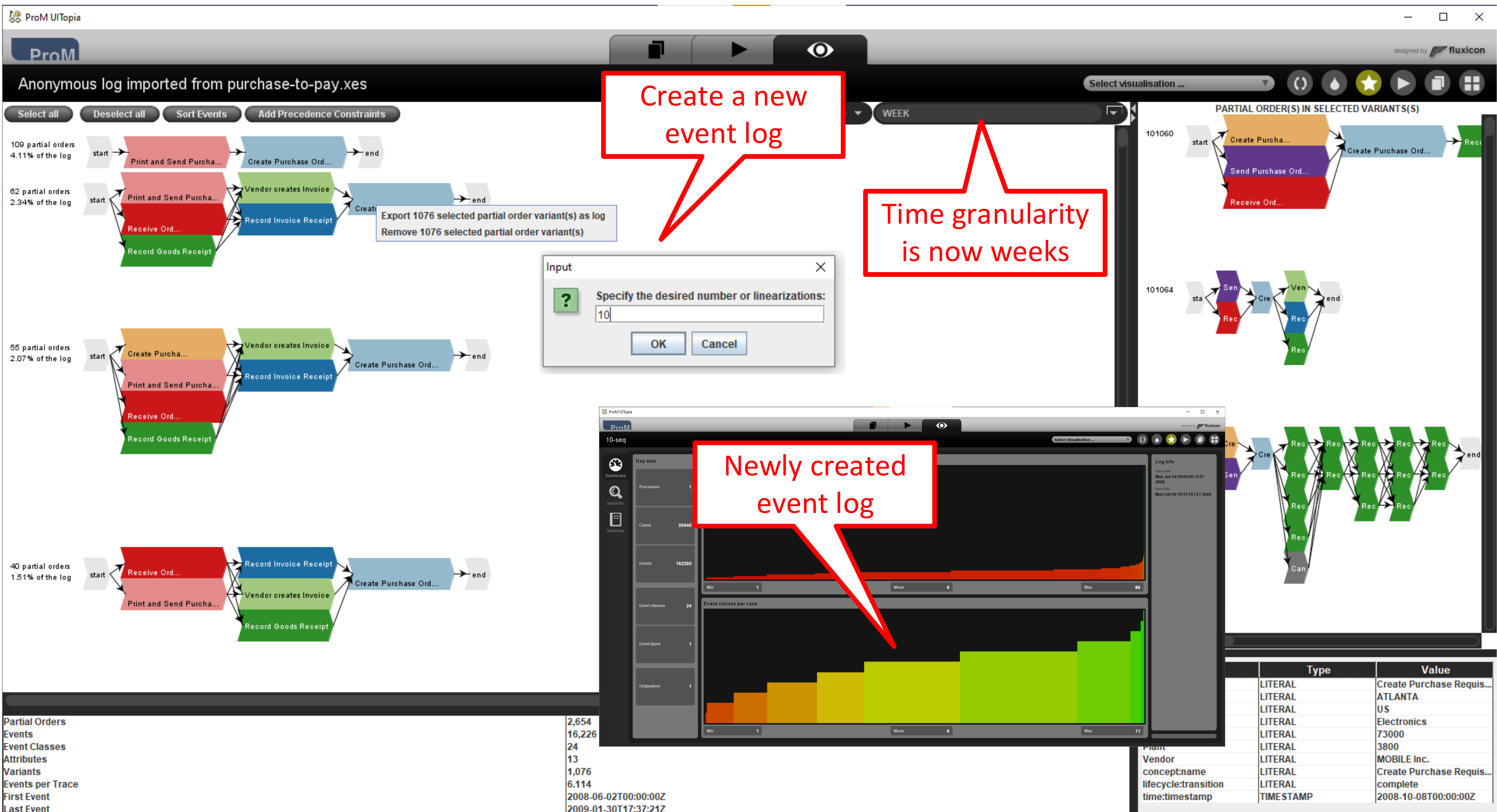}
\caption{Another view on the same P2P dataset now using a time granularity of a week. A new event log was created using a $k$-sequentialization (with $k=10$).}\label{fig-tool2}
}
\end{figure}

Figure~\ref{fig-tool2} shows the \emph{PartialOrderVisualizer} for the same event data, but now using a time granularity of a week.
One can see that more events become unordered, because they happen in the same week.
Using this view, we generated a new event log replicating each partially-ordered case 10 times. 
As expected, this new event log has 26,540 cases and 162,260 events.
Interestingly, there are now 8,864 trace variants.

The \emph{PartialOrderVisualizer} has been applied to a range of event logs, e.g., we have used it to analyze the treatment of Covid-19 patients at Uniklinik RWTH Aachen. In this Covid-19 dataset, only the dates are reliable. Na\"{\i}vely using the ordering in the event log or the recorded timestamps leads to incorrect conclusions.

For the Covid-19 dataset it takes just a few seconds.  For a larger data sets like the well-known road fines event log\footnote{Road Traffic Fine Management Process, 4TU.ResearchData, \url{https://doi.org/10.4121/uuid:270fd440-1057-4fb9-89a9-b699b47990f5}}, which has over 560.000 events and 150.000 cases, it takes around 10 seconds (using for example the day, hour, minute, and second abstractions).

What is interesting is that in many applications the number of partially-ordered variants temporarily goes up when coarsening the time granularity. However, by definition, the number of partially-ordered variants is the smallest when all events are mapped onto the same time period.

\section{Conclusion}
\label{sec:concl}

The contribution of this paper is twofold. On the one hand, we discussed the possible interplay between time and order using an event log definition where events have timestamps and may be ordered by some partial order. Since (rounded) timestamps provide a strict weak ordering, the combination is a partial order (provided that the event log is consistent). On the order hand, we described a new way to preprocess event logs using 
a time aggregator $\mi{ta}$ and a tiebreaker $\prec_{tb}$.
The time aggregator makes the timestamps more coarse-grained to avoid accidental event ordering due to imprecise or partially incorrect timestamps.
The tiebreaker can be used to order events that have the same coarse-grained timestamp (e.g., date). The preprocessed event log can be created and explored using
the new \emph{PartialOrderVisualizer} implemented in ProM.
Viewing the event log at different time scales provides novel insights
and helps to counteract data quality problems.
Moreover, the \emph{PartialOrderVisualizer} can also generate a $k$-sequentialization of the partially-ordered event log. This allows for the application of regular process mining techniques (e.g., discovery and conformance checking).

Although some process mining techniques have been developed for partially-ordered event logs, we feel that more research is needed. Partial orders may be the result of uncertainty or explicit causal information. These need to be treated differently.
We also plan to integrate frequent item-set mining into our approach. The fact that certain combinations of activities often happen in the same period can be used to create event logs where high-level events refer to sets of low-level events.

\subsubsection*{Acknowledgments}
We thank the Alexander von Humboldt (AvH) Stiftung and the NHR Center for Computational Engineering Sciences (NHR4CES) for supporting our research.

\bibliographystyle{plain}
\bibliography{lit}

\end{document}